\documentclass[11pt,twoside]{article}
\usepackage{asp2010}
\usepackage{color}
\usepackage{natbib}
\usepackage{url}
\resetcounters

\begin{document}

\title{Panchromatic Calibration of Astronomical Observations \\
       with State-of-the-Art White Dwarf Model Atmospheres}

\author{Thomas Rauch}

\affil{Institute for Astronomy and Astrophysics,
       Kepler Center for Astro and Particle Physics,
       Eberhard Karls University, 
       Sand 1,
       72076 T\"ubingen, 
       Germany,
       rauch@astro.uni-tuebingen.de}

\begin{abstract}
Theoretical spectral energy distributions (SEDs)
of white dwarfs provide a powerful tool for
cross-calibration and sensitivity control
of instruments from the far infrared to the X-ray 
energy range.
Such SEDs can be calculated from
fully metal-line blanketed NLTE model-atmo\-spheres
that are e.g\@. computed by the
T\"ubingen NLTE Model-Atmo\-sphere Package (\emph{TMAP})
that has arrived at a high level of sophistication. 
\emph{TMAP} was successfully employed for the reliable
spectral analysis of many hot, compact post-AGB stars.

High-quality stellar spectra obtained over a wide energy
range establish a data base with a large number of 
spectral lines of many successive ions of different 
species. Their analysis allows to determine effective 
temperatures, surface gravities, and element abundances 
of individual (pre-)white dwarfs with very small error 
ranges.
We present applications of \emph{TMAP} SEDs for spectral 
analyses of hot, compact stars in the parameter range 
from (pre-) white dwarfs to neutron stars and 
demonstrate the improvement of flux calibration using 
white-dwarf SEDs that are e.g\@. available via registered
services in the Virtual Observatory. \\
~\\
\end{abstract}

\section{Introduction}
\label{sect:intro}

In an ideal case, the problem of flux calibration can be simplified to

\begin{equation}
\label{eg:basic}
f_\lambda = \frac{A}{d^2} F_\lambda\ ,
\end{equation}

\noindent
where $f_\lambda$ is the measured stellar flux, 
$A$ is the radiation area,
$d$ is the distance,
and $F_\lambda$ is the emitted flux.
In an astrophysical case, additional absorption in the
circumstellar,
interstellar, and Earth's atmosphere environment will
reduce $f_\lambda$. This absorption occurs on both,
time- and energy-dependent scales, and thus exacerbates
the calibration problem (cf\@. these proceedings).

While observational techniques have strongly improved in the 
last decades, and the measurement of $f_\lambda$ is
performed at 
high precision, 
high resolution, and 
high S/N
based on ground-based and space-borne instruments,
the main uncertainties of Eq.\,\ref{eg:basic} are found
on its right side. 
The first challenge to minimize these 
is to find ``simple'' objects with high photometric stability
whose radiation can be described by fundamental physics. 
White dwarfs (WDs) fulfill such requirements. 
Since their radius $R$
is determined be electron degeneracy, their radiative area can
be calculated from

\begin{equation}
\label{eg:area}
A = \pi R^2
\end{equation}

\noindent
with

\begin{equation}
\label{eg:area}
R = \sqrt{\frac{G\cdot M}{g}}
\end{equation}

\noindent
($M$ is the star's mass and $g$ its surface gravity.
$G$ is the gravitational constant.).

Most of the hot, hydrogen-rich WDs (spectral type DA)
with $T_\mathrm{eff} < 40\,000\,\mathrm{K}$ 
have virtually
pure hydrogen atmospheres due to gravitational settling
($\log g\,\raisebox{-0.10em}{$\stackrel{>}{{\mbox{\tiny $\sim$}}}$}\,7$),
while the hotter WDs exhibit lines of heavier elements
due to radiative levitation or weak stellar winds.
This singled out in the past stars with many metal lines in their spectra 
as standards for flux calibration.

Moreover, WDs are nearby and, thus, their distances $d$ can be measured
with high accuracy by parallaxes, particularly with the upcoming
\emph{GAIA}\footnote{\url{http://www.rssd.esa.int/index.php?project=GAIA&page=index}} mission.
A further advantage of their small distances is that WDs with extremely
low interstellar reddening can be found, where the calibration problem
reduces almost to the ideal case of Eq.\,\ref{eg:basic} -- at least for
space-borne missions. The spectra of such objects can be used for
panchromatic calibration (from the soft X-ray to the far infrared).

In this paper, we start with a brief summary on the
application of theoretical WD spectra to the flux calibration (Sect.\,\ref{sect:da}).
Then (Sect.\,\ref{sect:models}), we describe modeling
of WD atmospheres and the calculation of synthetic spectra.
In the case of G\,191-B2B is demonstrated that state-of-the-art 
metal-line blanketed model atmospheres reproduce well its spectrum (Sect.\,\ref{sect:g191b2b}).
In Sect.\,\ref{sect:cross}, we show example for the application
of DA model spectra in the cross-calibration between different
instruments,
An easy way to access such spectra is shown in Sect.\,\ref{sect:vo}.
We conclude in Sect.\,\ref{sect:conclusions}.

\section{Flux calibration using DA-type white dwarfs}
\label{sect:da}

Most of the presently known ($\approx 10\,000$) WDs are of spectral
type DA. Only a few of them are relatively bright and exhibit negligible reddening.
These are a priori suited to act as primary standard stars for the
flux calibration in a large variety of observatories.
As a pre-requisite, however, reliable model atmospheres are necessary to compare
observed with theoretical spectra. These became available in the early
1970ies, following the pioneering work of 
\citet{auermihalas1969,auermihalas1970} and \citet{mihalasauer1970}
who presented the first NLTE models for pure hydrogen 
and hydrogen + helium stellar atmospheres, respectively.
Consequently, \citet{oke1974} presented a compilation of 
absolute spectral energy distributions (SEDs) for 38 WDs
in the wavelength range $\lambda \ge 3200\,\mathrm{\AA}$, partly up to
more than 10\,000\,\AA.

\begin{figure}[ht!]
\plotone{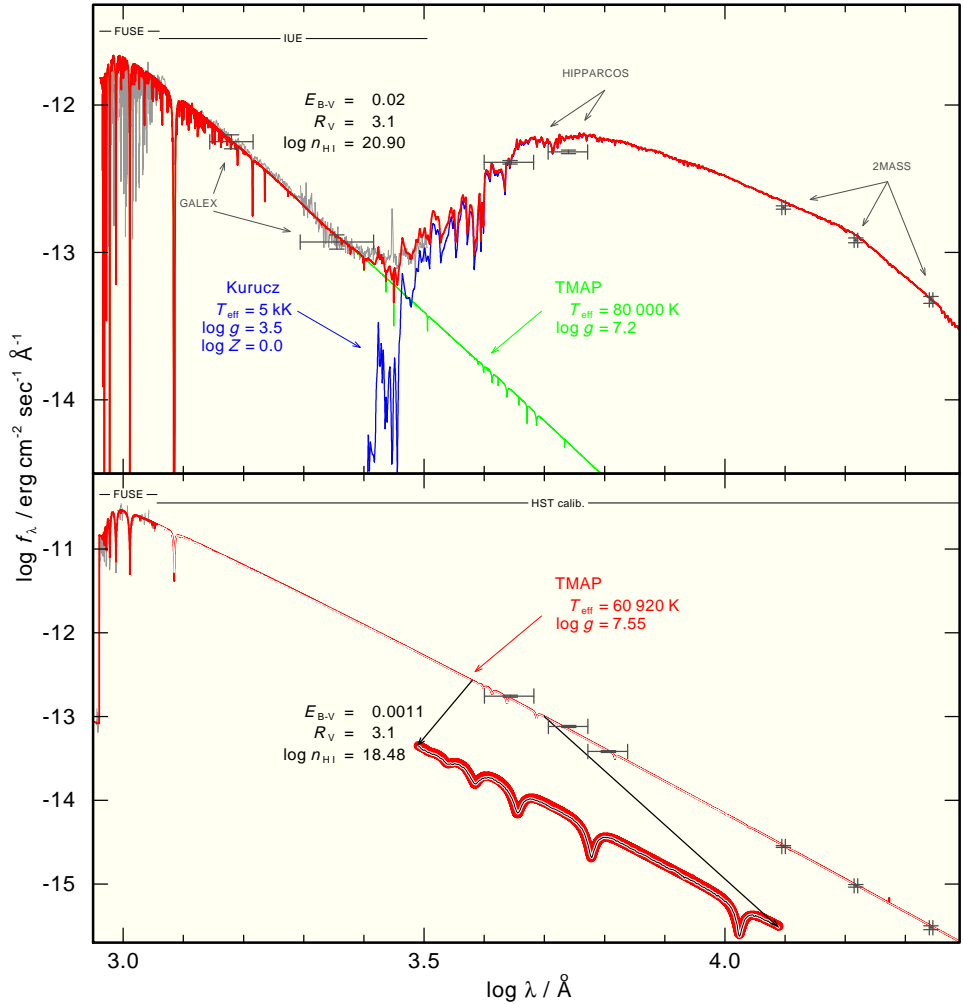}
\caption{Synthetic fluxes of an example binary (A\,35, top) and
         an isolated WD (G\,191-B2B, bottom) compared with observations.
         In case of A\,35, the composite model spectrum is normalized 
         to the 2MASS brightness at 22\,000\,\AA.
         The synthetic spectrum for G\,191-B2B is normalized to that of the   
         calibrated HST spectrum (R\@. Bohlin,
         priv\@. comm.). Since the observed and theoretical flux levels are 
         almost identical, the observation is plotted in white in the
         foreground of the  \emph{TMAP} spectrum (red) for clarity (visible
         in the electronic version).
         The insert shows a scaled-up part of the spectrum.
         Interstellar reddening was considered using the law of \citet{fitzpatrick1999}.
}
\label{fig:reddening}
\end{figure}

Figure\,\ref{fig:reddening} illustrates the advantage of 
an isolated DA WD in comparison to a DAO WD in a binary system -- 
the many strong metal lines in the DAO spectrum hamper an accurate calibration
in the FUV and the contribution of a cool companion spoils the 
low-energy region. Although the composite synthetic spectrum matches
the FUV flux as well as the optical and IR magnitudes, the uncertainties
in modeling both components cannot be quantified properly. This excludes
such objects from the standards of \emph{panchromatic} flux calibration.
An isolated DA WD allows to normalize observation and synthetic spectrum
to the WD brightness in the infrared where interstellar reddening is lowest.
Note that there are no strong interstellar lines in the \emph{FUSE}
spectrum of G\,191-B2B (Fig.\,\ref{fig:reddening}).

Oke's list included HZ\,43A 
\citep[WD\,1314+293, for the WD numbers cf\@.]
[and their online catalogue\footnote{\url{http://www.astronomy.villanova.edu/WDCatalog/index.html}}]
{mccooksion1999} and 
G\,191-B2B (WD\,0501+527).
From a black body comparison, Oke found that the observed SED of G\,191-B2B corresponds
to a black body at $T = 100\,000\,\mathrm{K}$. In addition, HZ\,43A and G\,191-B2B 
were amongst the brightest of Oke's list. Together with
GD\,71 (WD\,0549+158) and
GD\,153 (WD\,1254+223), these were used by \citet{bohlinetal1995} as
primary standard stars for the Hubble Space Telescope (HST, Tab.\,\ref{tab:hst}).

\begin{table}[ht!]\centering
\caption{Parameters of the HST DA standard stars used by \citet{bohlinetal1995}.
         The $m_\mathrm{B}$ and $m_\mathrm{V}$ values are from SIMBAD.}
\label{tab:hst}
\begin{tabular}{llccccl}
\hline
\noalign{\smallskip}
& 
&
$T_\mathrm{eff}$ & 
$\log g$ & 
&
&
\vspace{-2mm}\\
WD name & 
name &
& 
& 
$m_\mathrm{B}$ &
$m_\mathrm{V}$ &
comment \vspace{-2mm}\\
& 
&
[K] & 
[$\mathrm{cm/sec^2})$] & 
&
&
\\
\noalign{\smallskip}
\hline
\noalign{\smallskip}
WD\,0549+158 & GD\,71   & 32\,300 & 7.73 & 12.78 & 13.03 & \\
WD\,1254+223 & GD\,153  & 38\,500 & 7.67 & 13.17 & 13.40 & \\ 
WD\,0501+527 & G191-B2B & 61\,300 & 7.50 & 11.44 & 11.69 & metal lines \\
WD\,1314+293 & HZ\,43A  & 50\,000 & 8.00 & 12.52 & 12.66 & binary \\
\hline
\end{tabular}
\end{table}

The HST flux scale \citep[1200\,\AA\ - 8000\,\AA]{bohlinetal1995}
was the reference for other calibration work, e.g\@.
\citet{holbergbergeron2006} used it to calibrate four major ground-based
systems 
(Johnson-Kron-Cousins UBVRI, 
Str\"omgren uvby,
2MASS\footnote{\url{http://www.ipac.caltech.edu/2mass/}} JHK$_\mathrm{s}$,
SDSS\footnote{\url{http://www.sdss.org/}} ugriz).

Although stellar model atmospheres improved continuously in the last four deca\-des,
some problems are still unsolved, e.g\@. the discrepancy between $T_\mathrm{eff}$
determined from H\,{\sc i} lines in ultraviolet and optical spectra \citep{barstowetal2001}.
In the case of G191-B2B, \citet{lajoiebergeron2007} found
$T^\mathrm{UV}_\mathrm{eff} = 60\,680\,\pm\,15\,000\,\mathrm{K}$ and
$T^\mathrm{opt}_\mathrm{eff} = 57\,414\,\pm\,4\,700\,\mathrm{K}$.
Even improved Stark line-broadening tables for H\,{\sc i} Lyman and Balmer lines
\citep{tremblaybergeron2009}
leave this discrepancy unexplained.

The recent, comprehensive analysis of more than 1100 WDs \citep{gianninasetal2011}
shows significant deviations for the HST DA standard stars (Tab.\,\ref{tab:gianninas})
from previous values (Tab.\,\ref{tab:hst}). Note that the errors given are from their
statistical approach and, thus, appear too optimistic. A calibration of \emph{IRAC}
aboard \emph{Spitzer ST} \citep{bohlinetal2011} uses a $T_\mathrm{eff} = 61\,196\,\mathrm{K}$
model for G191-B2B which is within the error limits of \citet{gianninasetal2011} and fulfils
the desired 1\,\% flux-accuracy limit.

\begin{table}[ht!]\centering
\caption{Parameters of the HST DA standard stars \citep{gianninasetal2011}.}
\label{tab:gianninas}
\begin{tabular}{lr@{\,$\pm$\,}lr@{\,$\pm$\,}l}
\hline
\noalign{\smallskip}
& 
\multicolumn{2}{c}{$T_\mathrm{eff}$} & 
\multicolumn{2}{c}{$\log g$} \vspace{-2mm}\\
name & 
\multicolumn{2}{c}{}& 
\multicolumn{2}{c}{}\vspace{-2mm}\\
& 
\multicolumn{2}{c}{[K]} & 
\multicolumn{2}{c}{[$\mathrm{cm/sec^2}$]} \\
\noalign{\smallskip}
\hline
\noalign{\smallskip}
GD\,71           &  33\,590 &  \hbox{}\hspace{2mm}483 & \hbox{}\hspace{4mm}7.93 & 0.05     \\
GD\,153          &  40\,320 &  \hbox{}\hspace{2mm}626 & 7.93 & 0.05                        \\ 
G191-B2B         &  60\,920 &  \hbox{}\hspace{2mm}993 & 7.55 & 0.05                        \\
HZ\,43A          &  51\,116 &                    1249 & 7.90 & 0.07                        \\
\hline
\end{tabular}
\end{table}

\section{Model atmospheres}
\label{sect:models}

\begin{figure}[ht!]
\plotone{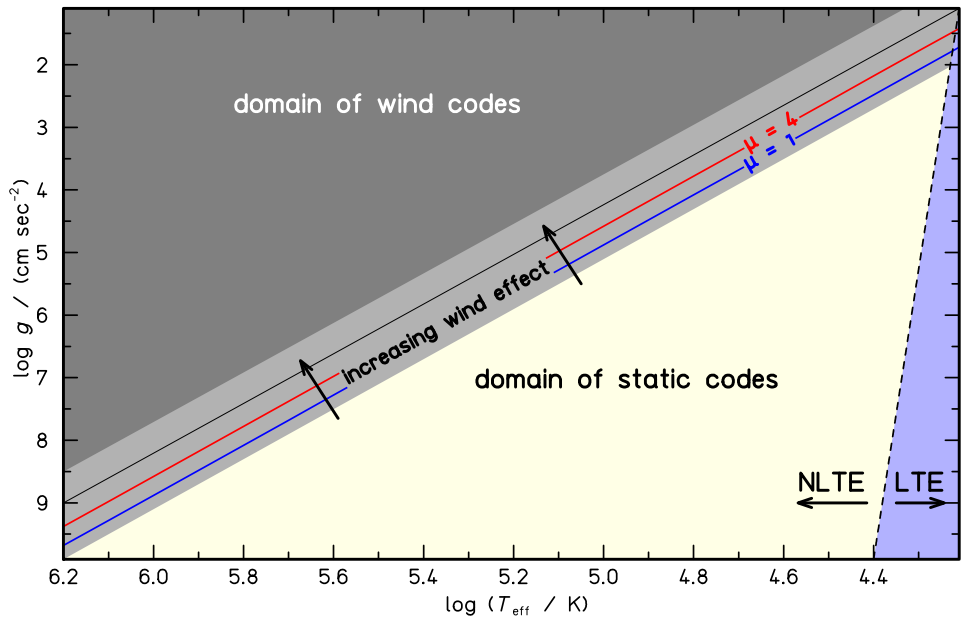}
\caption{Reliability domains of LTE and NLTE models for
         static and expanding stellar atmospheres.}
\label{fig:ltenlte}
\end{figure}

WD spectral analysis requires adequate observations (WDs are intrinsically faint)
and state-of-the-art theoretical atmosphere models \citep[cf\@.][]{oke1974} that account for
reliable physics and deviations from the assumption of a local
thermodynamic equilibrium (LTE). Figure\,\ref{fig:ltenlte} shows
that for cool stars with high surface gravity (spectral type B and later),
LTE model atmospheres might be sufficient \citep{auermihalas1972}. 
There are, however, NLTE effects
in any star, at least towards higher energies and in high-resolution
spectra. Aiming at a panchromatic flux calibration 
(at an 1\,\% flux-error limit, which was the designated aim mentioned in many talks
at this conference) 
and a cross-correlation of various instruments from the soft X-ray to the infrared, 
the use of NLTE stellar-atmosphere models is mandatory.

Since reliable atomic data is a crucial input for model-atmosphere calculations,
we start with a brief summary of existing lacks and problems in
Sect.~\ref{sect:atomicdata}. Then, we demonstrate the successful
application of NLTE atmosphere models in detailed spectral analysis and
show recent discoveries (Sect.~\ref{sect:reliability}).

\subsection{Atomic Data}
\label{sect:atomicdata}

\begin{figure}[ht!]
\plotone{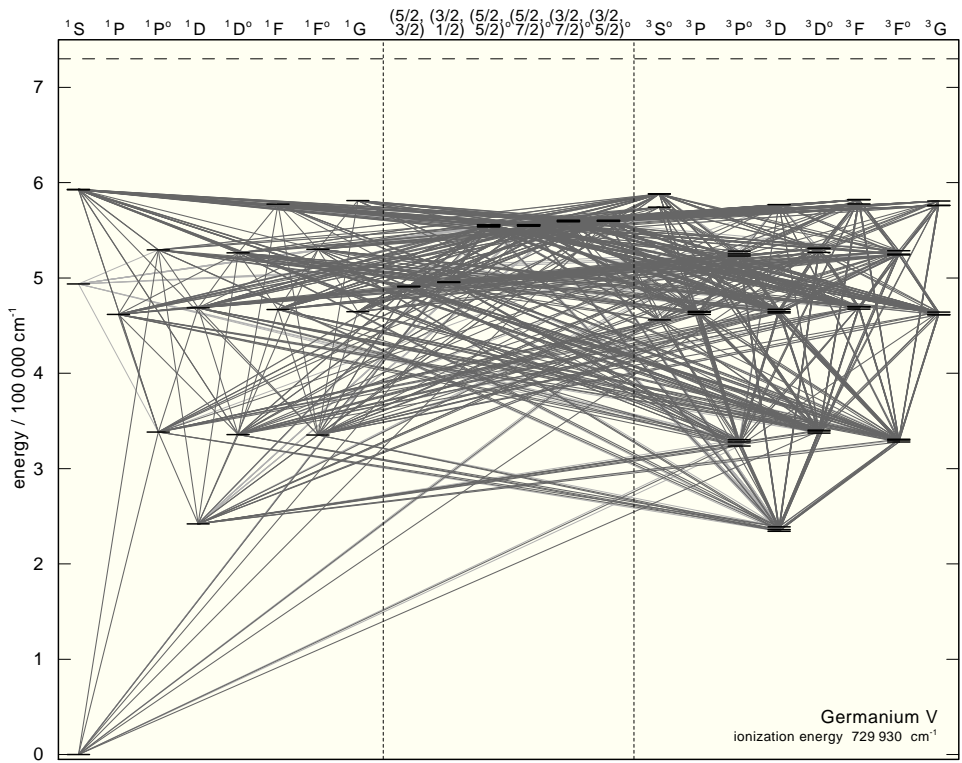}
\caption{Grotrian diagram of the \emph{TMAD} Ge\,{\sc v} model ion.}
\label{fig:grotrian}
\end{figure}

For \emph{TMAP}, we compile atomic data from several standard sources like
NIST\footnote{\url{http://www.nist.gov/pml/data/asd.cfm}},
Kelly's data\-base\footnote{\url{http://www.cfa.harvard.edu/ampcgi/kelly.pl}}, or
CHIANTI\footnote{\url{http://www.chianti.rl.ac.uk}},
the Opacity \citep{seatonetal1994} 
and 
Iron \citep{hummeretal1993}
projects\footnote{\url{http://cdsweb.u-strasbg.fr/topbase/home.html}}
as well as
Kurucz's line lists\footnote{\url{http://kurucz.harvard.edu/}} \citep{kurucz1991,kurucz2009}.
The model atoms \citep{rauchdeetjen2003} are provided by the
T\"ubingen Model-Atom Database \emph{TMAD}\footnote{\url{http://astro.uni-tuebingen.de/~TMAD}}
(Sect.~\ref{sect:vo}). \emph{TMAD} is continuously updated and extended for the most
recent atomic data. Presently it provides
ready-to-use, \emph{TMAP}-compliant model atoms of the elements H - Ca, Ge (Fig.~\ref{fig:grotrian}), Kr, and Xe.

\subsection{On the reliability of NLTE model atmospheres}
\label{sect:reliability}

\emph{TMAP}\footnote{\url{http://astro.uni-tuebingen.de/~TMAP}} \citep{werneretal2003}
can calculate fully metal-line blanketed WD model atmospheres.
About 1600 atomic levels and 7000 individual spectral lines of the
elements H - K can be treated in NLTE. Due to a statistical
treatment \citep{rauchdeetjen2003}, $\approx$\,300 million of lines of
the iron-group elements (Ca - Ni) can be considered. \emph{TMAP}
assumes hydrostatic and radiative equilibrium and a plane-parallel geometry.
It makes use of \emph{TMAD}.
\emph{TMAP} was successfully employed for detailed spectral analyses of hot, compact stars
\citep[e.g\@.][]{rauchetal2007,rauchetal2010,rauchetal2012,wassermannetal2010,werneretal2012}.
In these analyses, the models considered opacities of many elements from
H to Ni. 

\begin{figure}[ht!]
\plotone{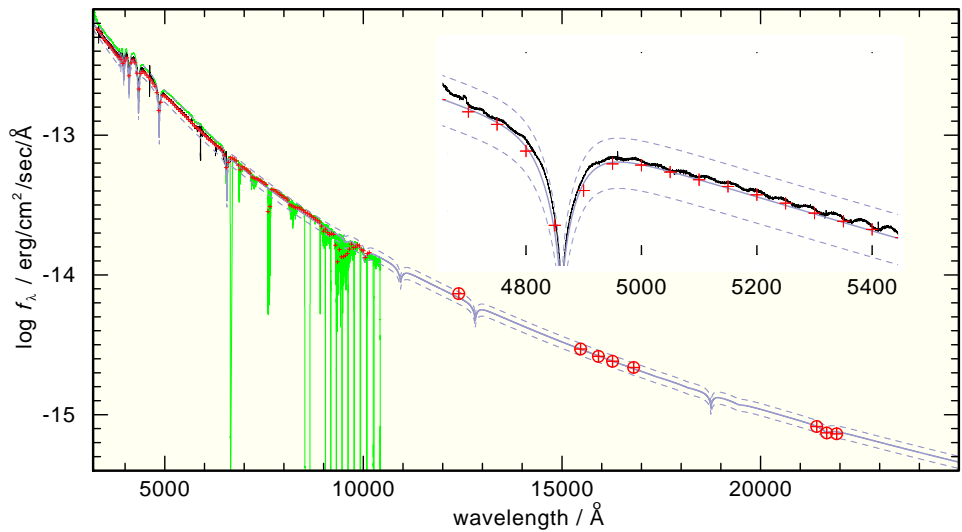}
\caption{Comparison of a \emph{TMAP} model (light blue / grey) for EG\,274 
($T_\mathrm{eff} = 24\,276\,\mathrm{K}$, $\log g = 8.01$)
with an optical spectrum and infrared brightness measurements.
The insert shows a detail of the UVES spectrum in the optical.
The dashed lines indicate the flux limits of the synthetic spectrum
within the $T_\mathrm{eff}$ and $\log g$ uncertainties.
}
\label{fig:vernet}
\end{figure}

In case of synthetic spectra for high-$g$ DA-type WDs (Sect.\,\ref{sect:intro}), 
where the calculation of pure hydrogen models is sufficient, the highest reliability
is achieved because H\,{\sc i} atomic data is accurately known.
Especially, tabulated Stark line-broadening tables are available \citep{lemke1997,tremblaybergeron2009}.
The models depend on $T_\mathrm{eff}$ and $\log g$ only, i.e\@. there are no other free parameters.
A synthetic WD spectrum can be calculated for a large wavelength range with
arbitrary resolution. Figure\,\ref{fig:vernet} shows the example of
EG\,274 (WD\,1620$-$391) for the optical to the near infrared wavelength range.
The observations were performed at ESO (European Southern Observatory, Chile)
with 
UVES (Ultraviolet and Visual Echelle Spectrograph)
and
SINFONI (Spectrograph for INtegral Field Observations in the Near Infrared) 
\citep[cf\@.][]{vernetetal2008a,vernetetal2008b}.
The initial aim to achieve an agreement of observation and theory better than 10\,\% in flux
was exceeded by far.

It is worthwhile to note that large observatories like ESO have large calibration
programs where spectro-photometric standard stars are observed every
night. The huge number of observations with different setups allows
to define master spectra (Fig.~\ref{fig:vernet}) that help to improve
the pipeline reduction of the observations.  These spectra may serve also
to automatically measure e.g\@. the nights' conditions or to monitor the efficiency 
of a telescope as a whole by comparison of the master flux level of a star
to that of an actual observation. Moreover, the pipeline reduction can
be improved, e.g\@. to remove residuals of \'echelle orders (see insert
in Fig.~\ref{fig:vernet}, EG\,274 is a spectrum
from the Advanced Data Products\footnote{\url{http://archive.eso.org/eso/eso_archive_adp.html}}).

The capability and reliability of a model-atmospheres code can be judged
from its application to spectral analysis over a wide wavelength range.
In general, synthetic spectra are more reliable in the low energy range
because towards higher energies, the consideration of metal opacities
(bound-bound as well as bound-free transitions) and the lack of atomic
input data (Sect.~\ref{sect:atomicdata}) thwarts any atmosphere code, how
sophisticated it may be. A simple look at the model-atom statistics is not sufficient.
\citet{kudritzki1976} used the complete linearization technique of
\citet{auermihalas1969} and allowed 18 atomic levels 
(5 H\,{\sc i}, 1 H\,{\sc ii}, 5 He\,{\sc i}, 6 He\,{\sc ii},  1 He\,{\sc iii}) 
to deviate from LTE and included six H\,{\sc i} line transitions in his so-called
\emph{classical models}. Together with 65 frequency points, he arrived at a system of 
86 equations to solve simultaneously and at the accuracy limit of the 32\,bit computers that
were available. Present models employ the approximated lambda operator (ALO) method
\citep{wernerhusfeld1985,werner1986,werner1989,wernerdreizler1999} and consider all elements from
H to Ni \citep{rauch1997,rauch2003} with
e.g\@. 686 levels in NLTE, 2417 individual lines, and millions of lines of the
iron group in the analysis of LS\,V\,+46$^\mathrm{o}$21 \citep{rauchetal2007}. 

The large number of levels and lines that can be considered in the models,
however, implies the difficulty to quantify e.g\@. uncertainties of the respective
atomic data (Sect.~\ref{sect:atomicdata}) and, thus, the synthetic spectra.
Detailed spectral analyses of high-quality spectra and the agreement between
model and observation are an important measure to estimate the models' flux accuracy.
We will now present some individual examples for spectral analyses with \emph{TMAP}
models of compact stars with different $T_\mathrm{eff}$.

\boldmath
\subsubsection{(Pre-) white dwarfs: $T_\mathrm{eff} \approx 100\,000\,\mathrm{K}$}
\label{sect:wd}
\unboldmath

The main field of our working group are post-AGB stars, with a focus on
hydrogen-deficient stellar evolution. The development of observation techniques
and the quality of the obtained spectra in the ultraviolet wavelength range
with e.g\@. 
\emph{FUSE} (Far Ultraviolet Spectroscopic Explorer)
or
\emph{HST/STIS} (Hubble Space Telescope / Space Telescope Imaging Spectrograph)
allowed the determination of stellar photospheric parameters like
$T_\mathrm{eff}$, $\log g$, and elemental abundances with unprecedented
accuracy \citep[see][and references therein]{wernerherwig2006}.
The identification of hitherto unidentified lines in \emph{FUSE} and
\emph{STIS} spectra and abundance analyses of many of the respective species
(e.g\@. 
Ne\,{\sc vii}     \citep{werneretal2004},
F\,{\sc vi}       \citep{wrk2005}, 
Ar\,{\sc vi}      \citep{rauchetal2007}, 
Ar\,{\sc vii}     \citep{wrk2007a}, 
Ne\,{\sc viii}    \citep{wrk2007b}, 
Ca\,{\sc x}       \citep{wrk2008}, and
Fe\,{\sc x}       \citep{wrk2010},
Ga\,{\sc v}, 
Ge\,{\sc v}, 
As\,{\sc v}, 
Se\,{\sc iv - vi}, 
Kr\,{\sc vi, vii},
Mb\,{\sc vi}, 
Sn\,{\sc iv - v},
Te\,{\sc vi},
I\,{\sc vi},
Xe\,{\sc vi, vii} \citep{werneretal2012})
were the direct consequence of
a continuous development of \emph{TMAP} and the available atomic data.
E.g\@. the identification of Ne\,{\sc viii} lines \citep{wrk2007b} followed
the publication of energy levels and spectral lines of Ne\,{\sc viii}
by \citet{kramidaetal2006}.

\begin{figure}[ht!]
\plotone{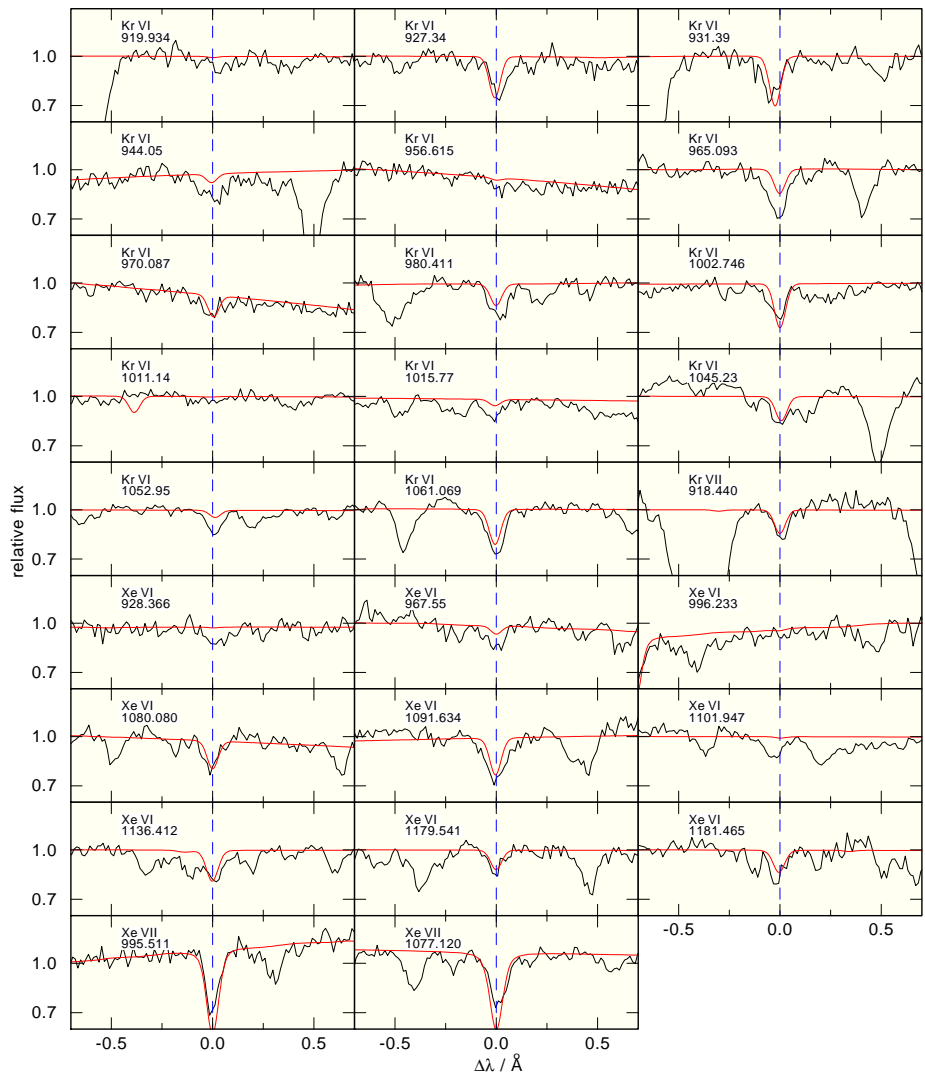}
\caption{Kr\,{\sc vi}, Kr\,{\sc vii}, Xe\,{\sc vi}, and Xe\,{\sc vii} lines
         in the FUV spectrum of RE\,0503$-$289.}
\label{fig:krxe}
\end{figure}

Figure~\ref{fig:krxe} shows newly identified Kr and Xe lines
in the \emph{FUSE} spectrum of the DO-type WD RE\,0503$-$289 \citep{werneretal2012}.
With these discoveries, new ionization equilibria, e.g\@. Kr\,{\sc vi} / {\sc vii},
can be evaluated for a more precise determination of $T_\mathrm{eff}$. In addition,
abundance determinations of trans-iron elements establish new challenges for
stellar evolutionary theory. Although \citet{werneretal2012} identified 
22 out of the 23 Ge\,{\sc v} lines that are listed at NIST and five that are
listed in Kelly's database, a Ge abundance analysis was impossible due to the lack of
transition probabilities. After their calculation, \citet{rauchetal2012} presented
a precise determination of the Ge abundance in RE\,0503$-$289 and discovered
additional four Ge\,{\sc iv}, 14 Ge\,{\sc v} and seven Ge\,{\sc vi} lines.

\begin{figure}[ht!]
\plotone{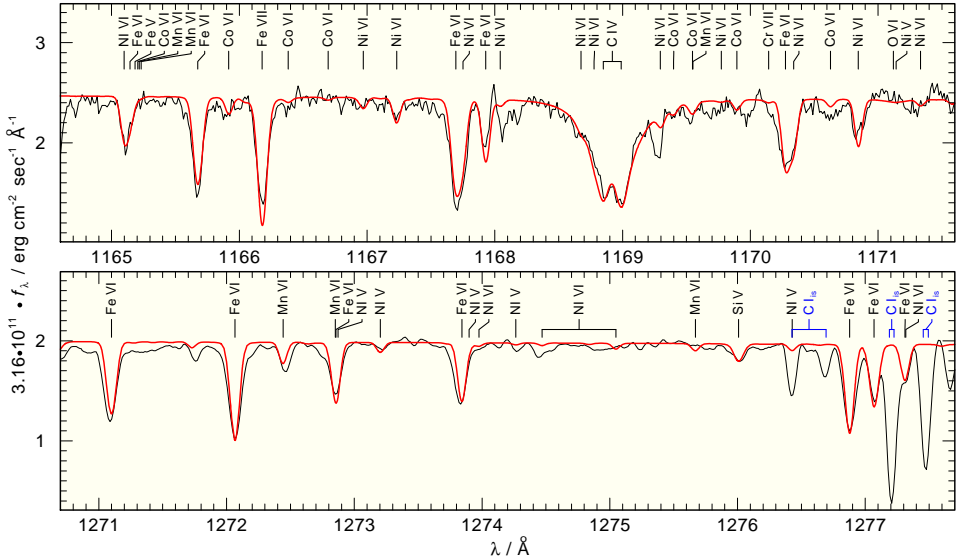}
\caption{Two sections of the 
         \emph{FUSE} (top) and
         \emph{STIS} (bottom) observations of LS\,V\,+46$^\mathrm{o}$21
         compared with the final model of \citet{rauchetal2007}.
         Photospheric lines are marked at top of the panels.
         is denotes interstellar origin (not modeled here).}
\label{fig:lsv}
\end{figure}

In the spectral analysis of LS\,V\,+46$^\mathrm{o}$21 \citep{rauchetal2007},
the central star of the planetary nebula Sh\,2$-$216,
it was possible to identify and reproduce about 95\% of all spectral
lines (more than 1000) in the \emph{FUSE} and \emph{STIS} spectra (Fig.~\ref{fig:lsv})
by the combination of a photospheric model and a model of the interstellar
line-absorption \emph{Owens} profile fitting procedure.

Figure~\ref{fig:lsv} demonstrates the unique opportunity to employ
stars as stellar laboratories for atomic physics. The large number of
spectral lines, e.g\@. of Fe\,{\sc vi}, allows to measure relative
line strengths and to verify or improve the respective $\log gf$ values
($g$ is here the statistical weight of the lower level of the line
transition, $f$ the oscillator strength). In addition,
precise wavelength measurements in high-resolution spectra allow to
improve the atomic level energies.

\boldmath
\subsubsection{Super-soft sources: $T_\mathrm{eff} \approx \mathrm{700\,000\,K}$}
\label{sect:sss}
\unboldmath

\begin{figure}[ht!]
\plotone{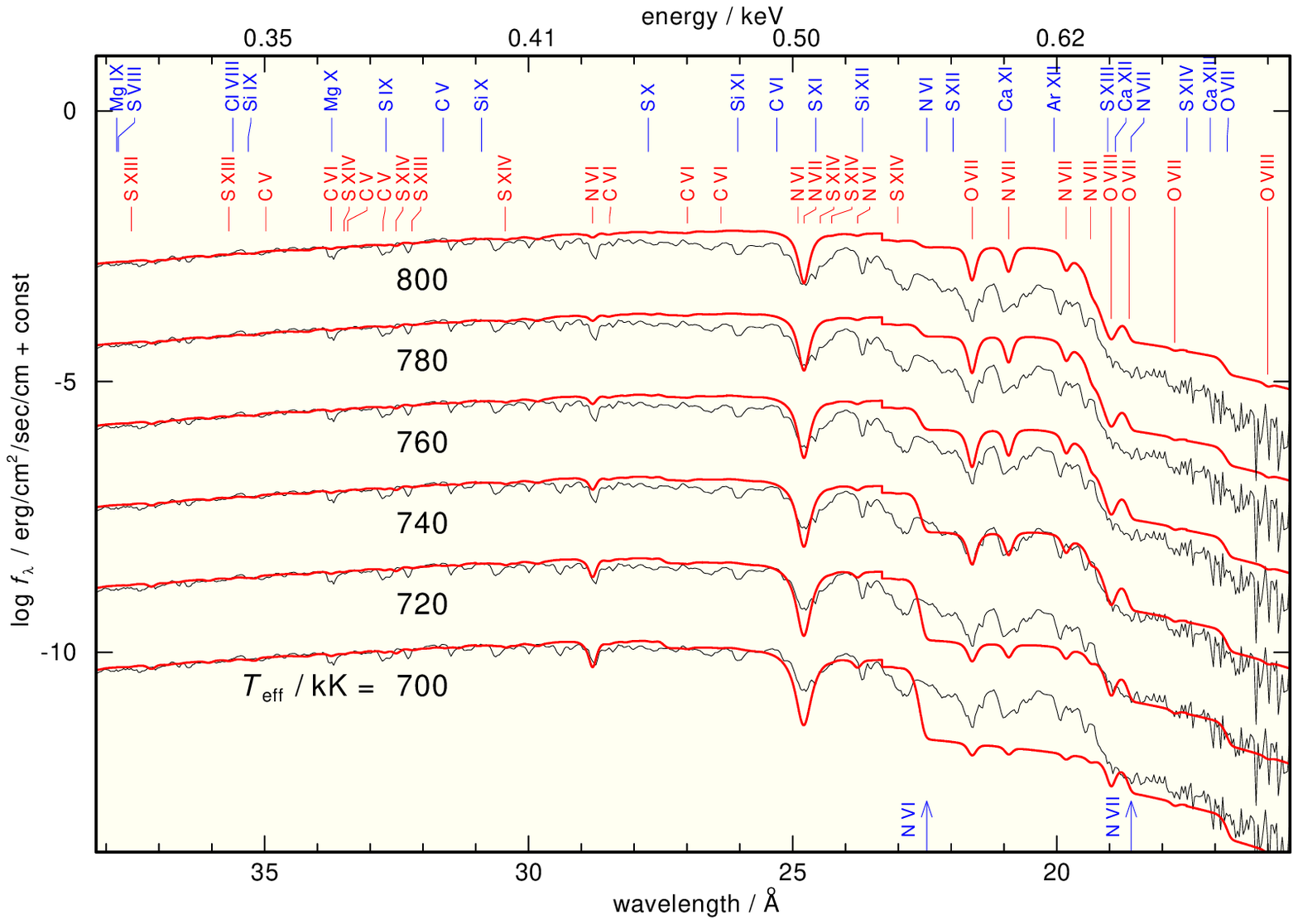}
\caption{Dependance of the strengths of lines (marked above the top spectrum, red)
         and ground-state thresholds (marked at top) on $T_\mathrm{eff}$.
         The SEDs of our models ($\log g = 9$) 
         are compared with the \emph{XMM-Newton} observation (2003-04-04)
         of V\,4743 Sgr and normalized to match the observed flux level at 38\,\AA.
         IG denotes a generic model atom that contains Ca - Ni.}
\label{fig:v4743}
\end{figure}

Some months after their outbursts, novae may evolve into very bright 
super-soft X-ray sources (SSS). V\,4743 Sgr (outburst Sep 20, 2002) 
was the brightest SSS
in March 2003 \citep{nessetal2003}. Although the origin and the
physics of the nova's spectrum is still under debate, \citet{rauchetal2010}
could reproduce well the strengths and shapes of lines and absorption edges
and determined 
$T_\mathrm{eff} = 740\,000 \pm 20\,000\,\mathrm{K}$. Fig.\,\ref{fig:v4743} shows
a comparison of theoretical spectra with an \emph{XMM-Newton} (X-ray Multi-Mirror Mission)
RGS (Reflection Grating Spectrometer) fluxed spectrum from 
\emph{BiRD}\footnote{\url{http://xmm.esac.esa.int/BiRD/}}.
Note that the given error is rather conservative, i.e\@. we arrive at an 
uncertainty level of better than 3\,\% even for these extremely hot objects.

\begin{figure}[ht!]
\plotone{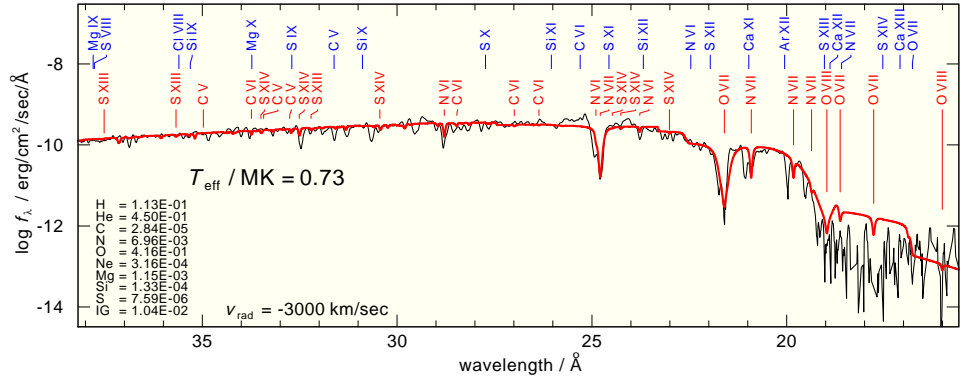}
\caption{SED of our preliminary \emph{TMAP} model  ($\log g = 9$) compared with 
         the \emph{Chandra} spectrum (2010-01-31) of 
         KT\,Eri (normalized to the observed flux at 38\,\AA).}
\label{fig:kteri}
\end{figure}

V\,4743 Sgr is not the only example where our static models were used successfully
to determine photospheric parameters of the WD in SSS. 
Nova KT \,Eri (outburst at Nov 25, 2009) was observed in early 2010 with
\emph{Chandra} and \emph{XMM-Newton}. 
In Fig.\ref{fig:kteri}, we show a
comparison of a model SED with the observation.
Although the model is not fine-tuned in abundances and does not include all
elements from H to Ni, the reproduction of the observed spectrum is already very good.
Obviously, lines of different ionization stages, e.g\@.  N\,{\sc vi} and N\,{\sc vii}
are formed in different radial velocity regimes but their theoretical strengths match
the observation.

\boldmath
\subsubsection{Neutron stars: $T_\mathrm{eff} \approx \mathrm{10\,000\,000\,K}$}
\label{sect:ns}
\unboldmath

At a typical $\log g \approx 14.39$ 
(neutron star with $M =1.4\,\mathrm{M_\odot}$ and  $R = 10\,\mathrm{km}$),
an extended Fig.~\ref{fig:ltenlte} would show that \emph{TMAP} models are still
valid up to $T_\mathrm{eff} \approx 20\,\mathrm{MK}$. At higher $T_\mathrm{eff}$, 
comptonization has a significant impact on the X-ray spectrum \citep{rauchetal2008}.

\begin{figure}[ht!]
\plotone{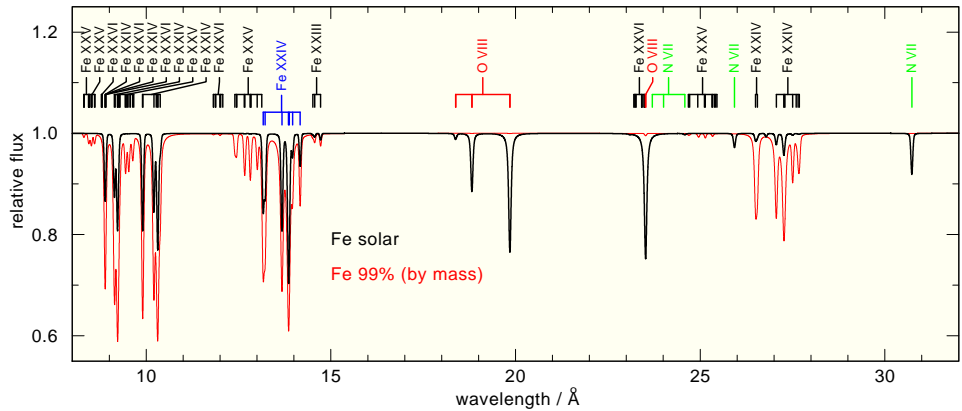}
\caption{Model spectra for EXO\,0748$-$676 with different iron content.}
\label{fig:exo}
\end{figure}

In the case of EXO\,0748$-$676, an X-ray bursting neutron star,
\citet{rauchetal2008} could easily show that the identification of 
gravitationally red-shifted absorption
lines of Fe\,{\sc xxv} and {\sc xxvi} by \citet{cottametal2002} is wrong because at the relevant temperatures,
Fe\,{\sc xxiv} lines should be much stronger but these were not observed (Fig.~\ref{fig:exo}).

\clearpage

\section{G\,191$-$B2B - too many metal lines?}
\label{sect:g191b2b}

G\,191$-$B2B is the visually brightest of the HST standard stars 
(Table~\ref{tab:gianninas}) and thus it may be the most observed
and best studied hot DA WD in all wavelength ranges. Hence, it should 
be the prime standard star for the UV wavelengths as well. However, 
it shows many weak metal lines \citep[e.g\@.][]{barstowetal2003}
in its observed UV spectrum. Therefore its was not used for the \emph{FUSE} calibration 
(G\@. Sonneborn, this conference). Detailed modeling with \emph{TMAP}
shows that these lines can be well reproduced and, thus, a reliable synthetic spectrum
from the H\,{\sc i} Lyman threshold to the far IR can be calculated
that allows a flux calibration over a wide energy interval.
Fig.~\ref{fig:reddening} shows a comparison of 
HST observations of 
G\,191$-$B2B\footnote{\url{http://www.stsci.edu/hst/observatory/cdbs/calspec.html}} 
within $1\,400 < \lambda < 30\,000\,\mathrm{\AA}$
with our preliminary \emph{TMAP} model that agree almost perfectly in the 
continuum-flux level. A detailed fine-tuning of the metal abundances is still ongoing 
that will improve the agreement in the FUV and EUV range.

\begin{figure}[ht!]
\plotone{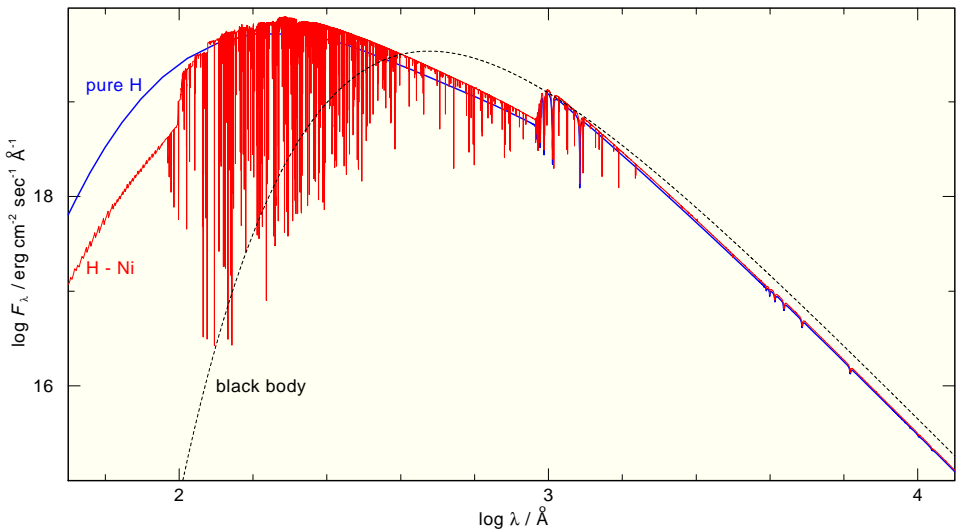}
\caption{Astrophysical flux of two \emph{TMAP} models for G\,191$-$B2B with different elemental
         compositions (blue, thick: pure H atmosphere; 
                       red, thin: H+He+C+N+O+Si+P+S+Ca+Sc+Ti+V+Cr+Mn+Fe+Co+Ni).
         The dashed line shows the black body flux at the same $T=60\,920\,\mathrm{K}$.}
\label{fig:g191flux}
\end{figure}

Figure~\ref{fig:g191flux} shows a comparison of the flux of a pure hydrogen model
to a metal-blanketed model and a black-body flux. Strong metal-line blanketing
is visible around the flux maximum in the H-Ni model. This results in significant 
deviations from the pure H model flux even in the IR. Compared to the model-atmosphere
fluxes, the black body has its flux maximum at a 2.7 times higher wavelength
and the peak maximum is about 40\% lower.

\section{Cross-correlation between various instruments}
\label{sect:cross}

\citet{beuermannetal2006} established two DA WDs, namely HZ43\,A and Sirius\,B,
and the neutron star RX\,J185635$-$3754 as soft X-ray standards. This enabled
a cross-calibration between the \emph{Chandra} 
LETG+HRC-S (Low Energy Transmission Grating, High Resolution Camera Spectroscopic array), the 
\emph{EUVE} (Extreme Ultraviolet Explorer) spectrometer, and the 
\emph{ROSAT} PSPC (R\"ontgensatellit, Position Sensitive Proportional Counter).
With their attempt, the effective area of \emph{ROSAT} could be corrected,
yielding consistent fluxes from about 20 to 160\,\AA.
It is worthwhile to note that during the work for this cross-correlation,
Jelle Kaastra discovered discrepancies between 
\emph{TMAP} and 
\emph{TLUSTY} \citep{hubenylanz1995} models in the case of HZ43\,A,
that were then intensively investigated and understood \citep[e.g\@.][]{rauch2008}.

\citet{drake2012} and the \emph{Chandra} LETG team relied on the growing confidence 
in the photospheric parameters for HZ43\,A \citep{beuermannetal2006}
and that its spectrum was well-represented by pure hydrogen atmospheric models. 
They used HZ43\,A spectra obtained over ten years and \emph{TMAP} atmospheres
in order to improve the in-flight calibration. The corrected effective area curve allowed
then to incorporate a secular decline in quantum efficiency for a better
(re-)calibration of LETG spectra. It is worthwhile to note that 
GD\,153 was newly introduced by 
\citet{menz2011} for the calibration of LETG spectra in order to bridge
the $T_\mathrm{eff}$ gap between Sirius\,B and HZ43\,A (Table~\ref{tab:standard}).

\section{Virtual Observatory Services and Tools for Synthetic Spectra}
\label{sect:vo}

\begin{figure}[ht!]
\plotone{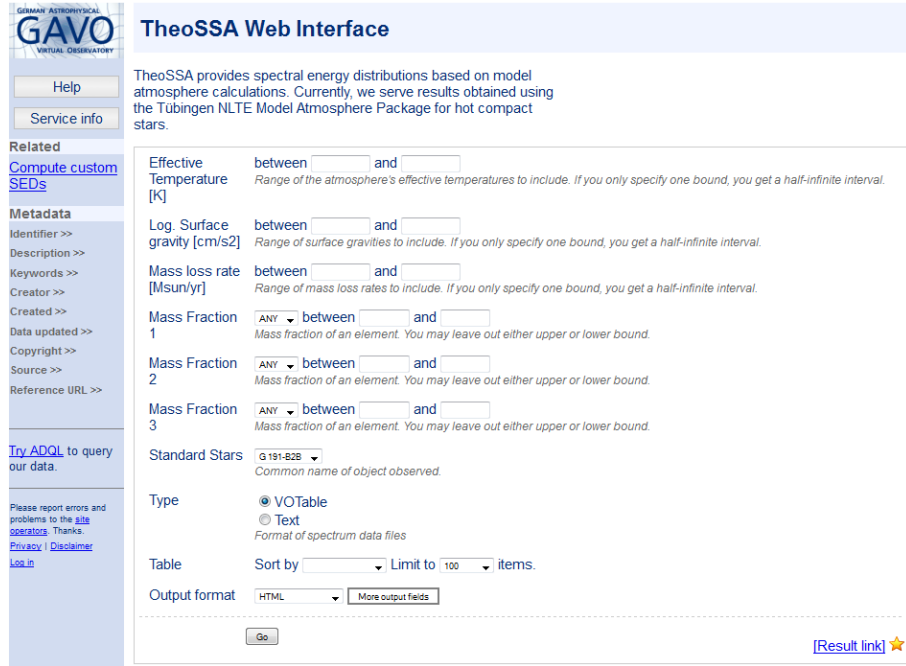}
\caption{\emph{TheoSSA} web interface at {\tt http://dc.g-vo.org/theossa}.}
\label{fig:theossa}
\end{figure}

In the framework of the \emph{Virtual Observatory}\footnote{\url{http://www.ivoa.net/}} (\emph{VO}),
the \emph{German Astrophysical Virtual Observatory}\footnote{\url{http://www.g-vo.org/}} (\emph{GAVO})
provides the registered \emph{VO} service 
\emph{TheoSSA} (Theoretical Spectra Simple Access,
Fig.\,\ref{fig:theossa}).
It was created to provide access to synthetic stellar SEDs at three levels.
1) Fast and easy download of precalculated SEDs by an unexperienced \emph{VO} user 
(no detailed knowledge of the model-atmosphere code necessary).
2) Calculation of SEDs (elements H, He, C, N, and O) with standard \emph{TMAD} model atoms 
for a preliminary analysis of individual objects via the
\emph{TMAP} web interface \emph{TMAW}\footnote{\url{http://astro.uni-tuebingen.de/~TMAW}}.
3) Advanced calculation of SEDs with especially tailored model atoms, e.g\@.
for a comparison with other stellar model-atmosphere codes.

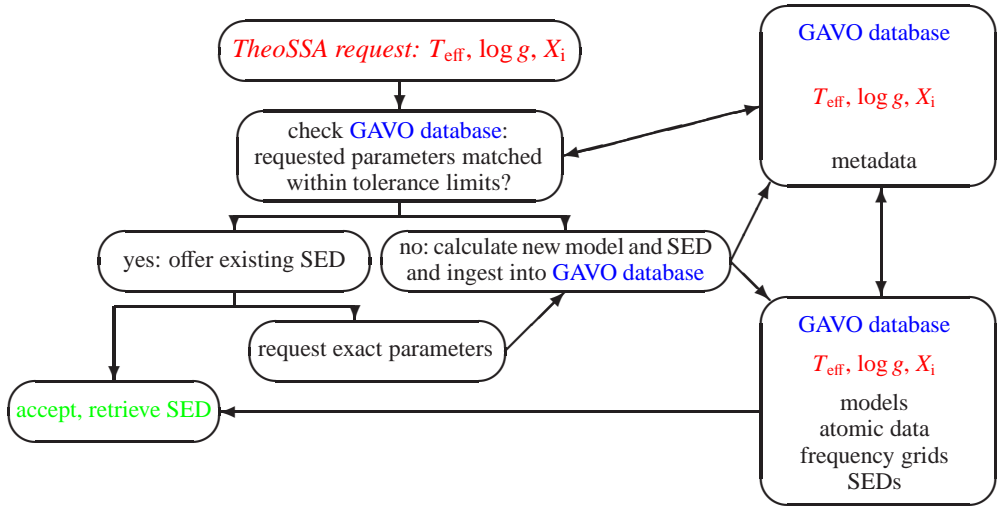
\begin{figure}[ht!]
\begin{picture}(17.0,10.0)
\thicklines

\put( 3.5, 8.0){\makebox(6.0,1.0)[c]{{\small\color{red}\emph{{TheoSSA} request:} $T_\mathrm{eff}$, $\log g$, $X_\mathrm{i}$}}}
\put( 4.0, 8.5){\oval( 1.0, 1.0)[l]}
\put( 9.0, 8.5){\oval( 1.0, 1.0)[r]}
\multiput( 4.0, 8.0)(0.0,1.0){2}{\line(1,0){5.0}}

\put( 6.5, 8.0){\vector(0,-1){0.5}}

\put( 4.54, 6.25){\makebox(4.0,1.0)[c]{{\begin{minipage}{60mm}
                                       \footnotesize
                                       \begin{center}
                                       check {\color{blue}GAVO database}:\\
                                       requested parameters matched\\
                                       within tolerance limits?
                                       \end{center}
                                       \end{minipage}
                                       }}}
\put( 4.3, 6.75){\oval( 1.0, 1.5)[l]}
\put( 8.7, 6.75){\oval( 1.0, 1.5)[r]}
\multiput( 4.3, 6.0)(0.0,1.5){2}{\line(1,0){4.4}}

\put( 6.5, 6.0){\line(0,-1){0.25}}
\put( 3.75, 5.75){\line(1,0){5.5}}
\multiput( 3.75, 5.75)(5.5,0.0){2}{\vector(0,-1){0.25}}

\multiput( 2.0, 5.0)(4.7,0.0){2}{\oval( 1.0, 1.0)[l]}
\multiput( 5.5, 5.0)(6.0,0.0){2}{\oval( 1.0, 1.0)[r]}
\multiput( 2.0, 4.5)(0.0,1.0){2}{\line(1,0){3.5}}
\multiput( 6.7, 4.5)(0.0,1.0){2}{\line(1,0){4.8}}
\put( 1.75, 4.5){\makebox(4.0,1.0)[c]{{\footnotesize yes: offer existing SED}}}
\put( 7.15, 4.5){\makebox(4.0,1.0)[c]{{\begin{minipage}{55mm}
                                       \footnotesize
                                       \begin{center}
                                       no:\,calculate\,new\,model\,and\,SED\\
                                       and ingest into {\color{blue}GAVO database}
                                       \end{center}
                                       \end{minipage}
                                       }}}

\put( 3.75, 4.5){\line(0,-1){0.25}}
\put( 1.75, 4.25){\line(1,0){4.0}}
\put( 1.75, 4.25){\vector(0,-1){1.25}}
\put( 5.75, 4.25){\vector(0,-1){0.25}}

\put( 0.5, 2.50){\oval( 1.0, 1.0)[l]}
\put( 3.0, 2.50){\oval( 1.0, 1.0)[r]}
\multiput( 0.5, 2.00)(0.0,1.0){2}{\line(1,0){2.5}}
\put( 0.25, 2.00){\makebox(3.0,1.0)[c]{{\footnotesize \color{green} accept, retrieve SED}}}
\put(12.50, 2.50){\vector(-1,0){9.00}}

\put( 4.5, 3.50){\oval( 1.0, 1.0)[l]}
\put( 7.75, 3.50){\oval( 1.0, 1.0)[r]}
\multiput( 4.5, 3.00)(0.0,1.0){2}{\line(1,0){3.25}}
\put( 3.85, 3.0){\makebox(4.5,1.0)[c]{{\footnotesize request exact parameters}}}
\put( 8.25, 3.5){\vector(1,1){1.0}}

\put(13.0, 7.75){\oval( 1.0, 3.0)[l]}
\put(15.9, 7.75){\oval( 1.0, 3.0)[r]}
\multiput(13.0, 6.25)(0.0,3.0){2}{\line(1,0){2.9}}
\put(12.7, 6.25){\makebox(3.5,3.0)[c]{{\begin{minipage}{40mm}
                                       \footnotesize
                                       \begin{center}
                                       {\color{blue}GAVO database}\vspace{5mm}\\
                                       {\color{red}$T_\mathrm{eff}$, $\log g$, $X_\mathrm{i}$}\vspace{5mm}\\
                                       metadata
                                       \end{center}
                                       \end{minipage}
                                       }}}

\put( 9.2, 6.75){\vector(4,1){3.30}}
\put(12.5, 7.58){\vector(-4,-1){3.30}}
\put(12.0, 5.00){\vector(1,2){0.653}}

\put(13.0, 2.675){\oval( 1.0, 3.45)[l]}
\put(15.9, 2.675){\oval( 1.0, 3.45)[r]}
\multiput(13.0, 0.95)(0.0,3.45){2}{\line(1,0){2.9}}
\put(12.7, 1.15){\makebox(3.5,3.0)[c]{{\begin{minipage}{50mm}
                                      \footnotesize
                                      \begin{center}
                                      {\color{blue}GAVO database}\vspace{1.7mm}\\
                                      {\color{red}$T_\mathrm{eff}$, $\log g$, $X_\mathrm{i}$}\vspace{1.7mm}\\
                                      models\\
                                      atomic data\\
                                      frequency grids\\
                                      SEDs
                                      \end{center}
                                      \end{minipage}
                                      }}}

\put(12.0, 5.00){\vector(1,-1){0.65}}

\put(14.5, 4.50){\vector(0,1){1.75}}
\put(14.5, 6.25){\vector(0,-1){1.85}}

\end{picture}
\caption{\emph{TheoSSA} / \emph{TMAW} data flow. $X_\mathrm{i}$ is the 
mass fraction of species $\mathrm{i} \in \left[\mathrm{H, He, C, N, O}\right]$.}
\label{fig:dataflow}
\end{figure}

\emph{TheoSSA} is related to the \emph{GAVO} database that includes already
thousands of SEDs. Newly calculated SEDs are automatically ingested and,
thus, this database is growing in time.
If a requested SED is not available, the \emph{VO} user will be redirected
to the \emph{TMAW} interface, where individual models can be calculated.
Figure\,\ref{fig:dataflow} illustrates the dataflow between the
\emph{TheoSSA} and \emph{TMAW} WWW interfaces and the \emph{GAVO} database.
For a pure H model, the calculation takes some hours, for H+He+C+N+O composed
models, one or two days are necessary. The wall-clock time, however, is
depending on the number of queued SED requests. 
The calculation of extended model grids ($> 150$ models) makes use of compute resources of 
\emph{AstroGrid-D\footnote{\url{http://www.gac-grid.de/}}}.

Presently, \emph{TheoSSA} provides a preliminary sample of pre-calculated DA WD spectra
(Table~\ref{tab:standard}) from the optical to infrared wavelength range. These
will be expanded to include the X-ray and UV ranges in the near future.

\begin{table}[ht!]\centering
\caption{Standard star SEDs ($3\,000 - 55\,000\,\mathrm{\AA}$) presently available in \emph{TheoSSA}.}
\label{tab:standard}
\begin{tabular}{lccl}
\hline
\noalign{\smallskip}
     & $T_\mathrm{eff}$ & $\log g$              &          \vspace{-2mm}\\
name &                  &                       & comments \vspace{-2mm}\\
     & [K]              & [$\mathrm{cm/sec^2}$] &          \\
\hline
\noalign{\smallskip}
EG\,274    & 24\,276 & 8.01 & \\
Feige\,67  & 75\,000 & 5.20 & \\
Feige\,110 & 40\,000 & 5.00 & He mass fraction 0.107 \\
GD\,108    & 22\,908 & 5.30 & \\
GD\,153    & 38\,205 & 7.89 & \\
GD\,50     & 40\,550 & 9.22 & \\
GD\,71     & 32\,780 & 7.83 & \\
G\,191-B2B & 60\,920 & 7.55 & metal-line blanketed model in preparation \\
G\,93-48   & 18\,100 & 7.85 & \\
HZ\,2      & 20\,600 & 7.90 & \\
HZ\,43     & 51\,116 & 7.90 & \\
Sirius\,B  & 24\,826 & 8.60 & \\
\hline
\end{tabular}
\vspace{3cm}
\end{table}

\section{Conclusions}
\label{sect:conclusions}

DA-type WDs are stable sources of radiation on a wide energy range
from the X-ray to the IR. This allows a cross-correlation between
ground- and space-based observations.

WD modeling arrived at a high level of sophistication and thus, 
they are well suited to provide model SEDs within 1\,\% flux-level accuracy.
It is, however,  difficult to quantify the absolute accuracy of model atmospheres,
at least if we aim at the ambitious 1\,\%.
About 200\,000\,000 variables are calculated iteratively to achieve convergence
of a NLTE stellar-atmosphere model.
Besides the mentioned uncertainties in atomic data and line-broadening theory,
systematic deviations may exist between different codes due to programming
language, coding, compiling, properties of the compute nodes, etc.
Future close collaboration of all groups working in stellar-atmosphere modeling 
is required \citep[in the legacy of past workshops,][]{beyond1991,sam2002} 
in order to identify and minimize existing uncertainties and to further improve 
the flux calibration which is a continuous process. The \emph{VO} provides an 
ideal platform to publish observations as well as models. The quality of spectral 
analysis and, hence, stellar-atmosphere models benefits from the observational data 
that it is based on.

Further improvement of flux calibration will be possible with \emph{GAIA}.
Precise distances of almost all known WDs will be available in the post-\emph{GAIA} era.

For ground-based telescopes, 
it is a major challenge to model and to correct for Earth's atmosphere 
contributions (cf\@. a variety of papers in these proceedings)
and to establish an extended database of secondary standard stars,
that should be observable in the vicinity of science targets.

\clearpage

\acknowledgements 
This work was supported by the German Aerospace Center (DLR, grant 05\,OR\,0806).  
I thank Klaus Werner for comments and reading the manu\-script.
This research has made use of the SIMBAD database, operated at CDS, Strasbourg, France.
Some of the data presented in this paper were obtained from the 
Mikulski Archive for Space Telescopes (MAST). STScI is operated by the 
Association of Universities for Research in Astronomy, Inc., under NASA 
contract NAS5-26555. Support for MAST for non-HST data is provided by 
the NASA Office of Space Science via grant NNX09AF08G and by other 
grants and contracts.

\bibliography{rauch_t}

\end{document}